\documentclass[aps,reprint,twocolumns,pre,superscriptaddress,floatfix,notitlepage,nofootinbib]{revtex4-2}
\usepackage{amssymb,amsmath,amsfonts}
\usepackage{mathtools}
\usepackage{physics}
\usepackage{enumerate}
\usepackage{array}
\usepackage{algorithm}
\usepackage{algpseudocode}
\usepackage{graphicx,epsfig,xcolor}
\usepackage[colorlinks=true,citecolor=blue,linkcolor=red]{hyperref}
\usepackage{newtxtext,newtxmath}

\begin{document}
\author{\'{A}lvaro Rubio-Garc\'{i}a}
  \email[]{alvaro.rubio@csic.es}
  \affiliation{Instituto de F\'{\i}sica Fundamental, IFF-CSIC, Serrano 113, 28006 Madrid, Spain}

\author{Juan Jos\'{e} Garc\'{i}a-Ripoll}
  \affiliation{Instituto de F\'{\i}sica Fundamental, IFF-CSIC, Serrano 113, 28006 Madrid, Spain}

\author{Diego Porras}
  \affiliation{Instituto de F\'{\i}sica Fundamental, IFF-CSIC, Serrano 113, 28006 Madrid, Spain}

\date{\today}

\title{Portfolio optimization with discrete simulated annealing}

\begin{abstract}
Portfolio optimization is an important process in finance that consists in finding the optimal asset allocation that maximizes expected returns while minimizing risk. When assets are allocated in discrete units, this is a combinatorial optimization problem that can be addressed by quantum and quantum-inspired algorithms. In this work we present an integer simulated annealing method to find optimal portfolios in the presence of discretized convex and non-convex cost functions. Our algorithm can deal with large size portfolios with hundreds of assets. We introduce a performance metric, the time to target, based on a lower bound to the cost function obtained with the continuous relaxation of the combinatorial optimization problem. This metric allows us to quantify the time required to achieve a solution with a given quality. We carry out numerical experiments and we benchmark the algorithm in two situations: (i) Monte Carlo instances are started at random, and (ii) the algorithm is warm-started with an initial instance close to the continuous relaxation of the problem. We find that in the case of warm-starting with convex cost functions, the time to target does not grow with the size of the optimization problem, so discretized versions of convex portfolio optimization problems are not hard to solve using classical resources. We have applied our method to the problem of re-balancing in the presence of non-convex transaction costs, and we have found that our algorithm can efficiently minimize those terms.
\end{abstract}

\maketitle

\section{Introduction}
\label{sec:intro}

Portfolio optimization is a task in the field of quantitative finance that consists in selecting the best distribution of assets according to some objectives, such as maximizing return and minimizing risk~\cite{Cornuejols2018}. Typically, portfolio optimization must include constraints such as a limited budget or requirements on investment diversification. This is a central problem in financial mathematics and it has been addressed by a large body of technical literature in finance and operational research. It is widely believed that portfolio optimization may pose a hard computational problem, specially in situations in which it becomes a non-convex optimization task. For that reason it is an ideal testbed for new methods and computational technologies, like optimization numerical methods or quantum computing.

This financial problem is often formulated using Markowitz's portfolio optimization theory~\cite{Markowitz1952,Kolm2014} (MPT). In its most basic form, this framework relies on the minimization of a quadratic utility, which aims to minimize the overall risk---measured by the portfolio's volatility---while maximizing the average returns. This utility function may be combined with further boundary conditions, such as bands that constrain the minimum and maximum weight of an asset in the portfolio. Both the original MPT and its banded version are both convex functions that can be efficiently optimized using quadratic programming methods, such as interior point methods.

More realistic models involve restrictions that turn MPT into a non-convex and thus a more difficult problem. One of such ingredients is the addition of transaction costs, non-convex functions describing the fees or taxes paid whenever a certain asset is purchased or sold. Fixed (non-convex) transaction costs are very relevant when considering periodic (monthly or even daily) re-balancing of a portfolio, as they may critically affect the average return of the strategy. Another non-convex constraint is the consideration of fixed-lot investments---e.g. stocks are bought in indivisible units, either one by one, or in predefined amounts---, leading to a discretization of the investment variables, possibly combined with a cardinality constraint on the total number of assets or total invested money, a situation that is particularly relevant in the case of small size portfolios. Several methods have been proposed to address such non-convex portfolio optimization problems. A common strategy is the use of branch-and-bound methods~\cite{Cornuejols2018}, but these scale exponentially in the portfolio size. Another approach is to approximate the non-convex portfolio problem by the solution of a similar convex function.

Both the practical relevance and its potential difficulty~\cite{Bernard2008} have made portfolio optimization problems a popular benchmark in the field of \textit{quantum finance}~\cite{Mugel2020b,Bouland2020,Egger2021}. In general, much of the work done has been focused on solving the unconstrained single period portfolio optimization problem~\cite{Rebentrost2018,Kerenidis2019,Gacon2020,Cohen2020,Cohen2020b,Cohen2020c,Hegade2021,Yalovetzky2021,Baker2022}, while some work has also been done with constraints of interest to the financial community, such as investment bands~\cite{Palmer2021,Certo2022} or cardinality constraints~\cite{Venturelli2019,Slate2020,Phillipson2020,Lorenzo2021}. A harder version of the problem, the multi-period portfolio optimization problem, has also been explored with linear~\cite{Mugel2020} and fixed transaction costs~\cite{Rosenberg2016,Hodson2019} and minimum asset holding periods~\cite{Mugel2021}.

%
%

All of these works rely on mapping the portfolio optimization problem into a Quadratic Unconstrained Binary Optimization (QUBO) problem, where binary variables are mapped to single qubits. From there on, these QUBO problems are solved using variational quantum algorithms, such as the Variational Quantum Eigensolver or the Quantum Approximate Optimization Algorithm~\cite{Cerezo2021}, and commercial quantum annealers~\cite{Hauke2020}. An important limitation of all these demonstrations lays in the number of available qubits. With present and near-term devices having 100's of qubits, and the QUBO mapping requiring several qubits per asset---e.g. see~\cite{Mugel2020,Palmer2021} for two alternative mappings---, quantum computers can only address small problems with no practical quantum advantage. One may overcome the size limitation with strategies such as the Hybrid D-Wave annealer~\cite{Venturelli2019,Certo2022,Mugel2020,Cohen2020,Cohen2020b,Cohen2020c,Palmer2021}, which combines quantum solutions of small problems with classical algorithms that combine those solutions to address larger baskets, such as the S\&P-500 Index. However, in this scenario, it remains an important question to determine (i) what is the actual hardness of the portfolio optimization problem and (ii) whether these problems merit the use of quantum computing methods, or may be addressed using improved classical methods. Indeed, a promising alternative is the use of physics-inspired (sometimes quantum-inspired) algorithms, such as Simulated Annealing (SA), Simulated Quantum Annealing~\cite{Martonak2004} or Matrix Product States, some of which have already been demonstrated in large practical applications~\cite{Palmer2021}.


In this work we introduce a simulated annealing~\cite{Kirkpatrick1983,Cerny1985} algorithm that is specifically designed for portfolio optimization. We characterize its efficiency and show that it can handle portfolio optimization problems with a large number of variables. Our method works with integer allocations and generic non-convex cost functions, overcoming the practical limitations of mappings to QUBO problems, such as approximations of the constraints~\cite{Mugel2020} or the use of ancillary variables~\cite{Hodson2019}.
In particular, we show the following results:
\begin{enumerate}[(i)]
    \item  We present an integer version of the SA algorithm with a novel scheme for the introduction of quadratic constraints in the cost function. This feature is crucial for dealing with portfolio optimization problems, where a constraint on the total portfolio budget must be typically imposed. We have also numerically derived optimal choices for the SA hyperparameters.
    \item We introduce a performance metric for our integer SA algorithm that is based on the existence of a continuous relaxation of the problem.
    If the continuous version has a convex cost function, it can be exactly solved with linear programming methods~\cite{Cornuejols2018}, yielding a lower bound for the cost function of the original integer portfolio optimization problem.
    We use this lower bound to characterize the quality of the solution obtained with the integer SA algorithm for system sizes for which and exact solution is not feasible.
    This approach allows us to define the time to target, which is a relaxed version of the more usual time to solution metric (see for example~\cite{Ronnow2014}).
    \item We calculate our performance metric when applying our integer SA to discretized versions of convex portfolio optimization problems. We find that, when the initial samples used as a starting configuration of the SA are drawn close to the continuous optimal solution, the time to target remains constant against the number of available assets and the total budget, both of which are parameters that determine the size of the configuration space.
    \item We apply our integer SA algorithm to the discretized non-convex portfolio optimization problem, where non-convexity comes from the inclusion of fixed transaction costs in the quadratic utility. In this case we cannot define any performance metric, as the non convexity of the continuous version of the problem does not allow us to find an optimal continuous solution in polynomial time. However, by analyzing the multi period portfolio optimization problem we are able to show the versatility of the SA algorithm and that it is sensible to the introduction of non convex terms in the quadratic utility.
\end{enumerate}

Our paper is organized as follows. In Sec.~\ref{sec:portfolio_optimization} we introduce the single period portfolio optimization problem with linear and fixed transaction costs and its discretized version when assets can only be bought in integer numbers and the total budget money is finite. There we define a metric to measure the quality of a given portfolio when the underlying problem is convex. Sec.~\ref{sec:dsa} introduces the discrete SA (DSA) algorithm, the necessity of having a scheme for the quadratic budget constraint and shows how to tune its hyperparameters to obtain optimal portfolios using a grid search algorithm. We use DSA in Sec.~\ref{sec:benchmarking} for solving convex portfolio optimization problems with large numbers of assets (50-433) and budgets $10^4-10^8\,\$$ and estimate the time to target when solving this problem. In Sec.~\ref{sec:non-convex} we show how DSA can incorporate non convex functions in the cost function by analyzing the multi period portfolio optimization problem. Finally, we draw conclusions from our work in Sec.~\ref{sec:conclusions}.


\section{The portfolio optimization problem}
\label{sec:portfolio_optimization}

Given a set of financial assets, their expected returns and the correlations between their prices, the Markowitz portfolio optimization problem consists on finding the optimal asset allocation that maximizes the expected returns while minimizing the risk, as estimated from the correlations between portfolio assets~\cite{Markowitz1952}. Let us define a portfolio $\vec{\omega}$ with $L$ assets using the percentages of money allocated to each asset $\omega_i \in \mathbb{R}$. Markowitz's portfolio optimization problem in its simplest form requires finding the weights that maximize the quadratic utility
\begin{equation}
    Q(\vec{\omega}) = \vec{\mu}^T \vec{\omega} - \frac{\lambda}{2} \vec{\omega}^T S \vec{\omega}.
    \label{eq:quadratic_utility}
\end{equation}
Here, $\vec{\mu}\in\mathbb{R}^L$ is the vector of expected returns, $S\in\mathbb{R}^{L\times L}$ the assets' price correlation matrix and $\lambda>0$ the so called risk tolerance factor, which sets a balances between the risk in the portfolio and its expected returns.

The asset weights $\omega_i$ are often subject to practical constraints such as: (a) the total budget must be allocated $\sum_{i} \omega_i = 1$ and (b) only long positions are allowed, $\omega_{i} \geq 0$~\cite{Kolm2014}, which are the ones we will use in this work. Other constraints of financial relevance that could be imposed are: setting investment bands $\omega_{\min} \leq \omega_i \leq \omega_{\max}$ that enforce portfolio diversification, setting specific bands for some sectors---e.g. finance, technology, healthcare---, fixing a target risk or setting a minimal expected return.

Because the allocation weights are real numbers and the negative quadratic utility $-Q(\vec{\omega})$ is a convex function, the portfolio optimization problem can be efficiently solved in polynomial time with the system's size $L$ using classical convex programming algorithms~\cite{Cornuejols2018}.

\subsection{Transaction costs}
\label{sec:transaction_costs}

Asset trading is usually accompanied by transaction fees, which might come from brokerage fees, takes, bid-ask spreads or other sources~\cite{Kolm2014}. In this work we will consider two of the most common transaction fees: (a) linear costs $t_l$, proportional to the absolute value of the traded amount, either buy or sell; and (b) fixed costs $t_f$, which must be paid whenever we trade an asset, independent of the traded amount. These costs are modelled by
\begin{equation}
    T_{c}(\vec{\omega},\vec{\omega}^0) = t_f \sum_{i=1}^{L} \big[ 1 - \delta(\omega_i^0 - \omega_i) \big] + t_l\, P_0 \sum_{i=1}^{L} |\omega_i^0 - \omega_i|,
    \label{eq:transaction_costs}
\end{equation}
with $\vec{\omega}^0$ the previous portfolio weights, $t_f \in \mathbf{R}_+$ a fixed price in dollars, $t_l \in [0,1)$ a cost percentage and $P_0$ the total budget (in dollars) before rebalancing. While linear transaction costs are represented by a convex function, fixed transaction costs are non-convex. Therefore, having or not fixed transaction costs will affect how easily one can find optimal portfolios using classical resources. To include the transaction costs in the portfolio optimization problem we consider them as a negative return over the total portfolio budget. We thus redefine the utility of a portfolio as the quadratic utility minus the rescaled transaction costs
\begin{equation}
    Q_{t}(\vec{\omega},\vec{\omega}^0) = \vec{\mu}^T \vec{\omega} -\frac{T_{c}(\vec{\omega},\vec{\omega}^0)}{P_0} - \frac{\lambda}{2} \vec{\omega}^T S \vec{\omega}.
    \label{eq:quadratic_utility_costs}
\end{equation}

\subsection{Discrete portfolio optimization problem}

In general, one can buy only an integer number of shares of every asset, $n_i \in \mathbb{Z}$, and the available budget $P_0$ to spend in assets is finite. In this case the allocation weights become quantized
\begin{equation}
    \omega_i = n_{i} \frac{p_i}{P_0},\quad 0 \leq n_i \leq n_i^{\max},
\end{equation}
with $n_i^{\max} = \lfloor P_0/p_i \rfloor$ the maximum allowed number of shares and $p_i$ the price of asset $i$. The quantization of the weights transforms the portfolio optimization problem from the optimization of a continuous function into a combinatorial optimization problem, which is NP-complete~\cite{Mansini1999} and cannot be guaranteed to be solved in polynomial time.

Because the asset weights are now discretized, the budget constraint, $\sum_{i}\omega_i = 1$, becomes harder to exactly satisfy. Therefore, we relax this constraint by defining a new cost function for the portfolio optimization problem with the budget constraint as a quadratic function
\begin{equation}
    C(\vec{\omega},\vec{\omega}^0) = Q_t(\vec{\omega},\vec{\omega}^0)
    - \lambda_B \left( \sum_{i=1}^L \omega_i - 1 \right)^2,
    \label{eq:cost_function}
\end{equation}
with $\lambda_B$ a Lagrange multiplier that controls the strength of the constraint.

\subsection{Quality of a portfolio}

For every possible portfolio $\vec{\omega}$ that satisfies $\sum_i \omega_i = 1$, its quadratic utility $Q_t(\vec{\omega},\vec{\omega}^0)$ has an upper bound given by an optimal continuous portfolio $\vec{\omega}_{\textrm{opt}}^c \in \mathbb{R}^{\otimes L}$. In this is optimal portfolio the number of shares of each asset $n_i$ is not restricted to integer numbers, so it is not a valid portfolio, but we can still consider it as the theoretical optimum and set it as a target for discrete portfolios. Therefore, we could define the quality of a given portfolio $\vec{\omega}$ as the difference $Q_t(\vec{\omega}_{\textrm{opt}}^c,\vec{\omega}^0) - Q_t(\vec{\omega},\vec{\omega}^0)$. However, as we mentioned above, in some cases it might be desirable to relax the budget constraint $\sum_i \omega_i \approx 1$, so $Q_t(\vec{\omega}_{\textrm{opt}}^c,\vec{\omega}^0)$ is not an upper bound for $Q_t(\vec{\omega},\vec{\omega}^0)$ any longer. Thus, we define the quality of a portfolio $\vec{\omega}$ as the distance
\begin{equation}
    d(\vec{\omega},\vec{\omega}^0) = |Q_t(\vec{\omega}_{\textrm{opt}}^c,\vec{\omega}^0) - Q_t(\vec{\omega},\vec{\omega}^0)|,
    \label{eq:distance_metric}
\end{equation}
for which lower distances represent a better quality of the portfolio. This definition remains valid only when the budget constraint is almost satisfied, which is the case of the optimal portfolios found by DSA.

\section{Discrete simulated annealing}
\label{sec:dsa}

To solve the combinatorial optimization problem of maximizing the cost function of Eq.~\eqref{eq:cost_function} with integer asset allocations we use a DSA algorithm. In its most simple form, this algorithm is similar to a Markov Chain Monte Carlo algorithm where the temperature is lowered at each step~\cite{Kirkpatrick1983}. In the version we use in this work,  DSA proposes at each step of the computation a random change $\delta \vec{\omega}$ in the number of shares of an asset $n_i \rightarrow n_i \pm 1$, where the asset $i$ and the sign of the change are drawn from an uniform distribution. Then, this change is accepted with a probability
\begin{equation}
    p_{\textrm{accept}} =
    \min\left\lbrace 1,\,\exp \big[\beta(s)\cdot \left(C(\vec{\omega}+\delta \vec{\omega})-C(\vec{\omega})\right) \big] \right\rbrace,
\end{equation}
with $C(\vec{\omega})$ the cost function~\eqref{eq:cost_function} of the portfolio $\vec{\omega}$ (we omit $\vec{\omega}^0$ for simplicity), $s$ the current step of the Markov chain and $\beta(s)$ the inverse temperature. Due to the stochastic nature of DSA, for each simulation we run $n$ independent DSA computations, where $n$ ranges from $10^4$ to $\sim 10^8$ in this work.

The DSA algorithm requires a set of hyperparameters that must be optimized to ensure finding optimal portfolios:
\begin{itemize}
    \item The number of steps $N_s$, which must be large enough so that DSA can explore a large area of the portfolio configuration space.

    \item The temperature scheme $\beta(s)$, which defines how the temperature changes along the Markov chain. While several options exist for the temperature scheme, we choose a linear scheme $\beta(s) = \beta(0) + m\cdot s$, with $m>0$, as we have observed it provides the best results. For this scheme the temperature is monotonically decreasing.

    To set the initial and final inverse temperatures in the temperature scheme we follow a similar approach as in~\cite{Isakov2015}. We define first the cost of changing one asset of the initial configuration
    \begin{equation}
        \Delta_{i,a} = C(\vec{\omega}(0) + a\cdot\delta \vec{\omega}_i)-C(\vec{\omega}(0)),
    \end{equation}
    with $\delta \vec{\omega}_i$ a vector with all 0 entries except for asset $i$, whose entry is $1$, and $a = \pm 1$. Then, we define the initial and final inverse temperatures for the cooling scheme as
    \begin{equation}
            \beta(0) = \frac{c_0}{\max_{i,a} \left\lbrace \big| \Delta_{i,a} \big|\right\rbrace},\quad
            \beta(N_s) = \frac{c_N}{\min_{i,a} \left\lbrace \big| \Delta_{i,a} \big|\right\rbrace },
        \label{eq:cooling_scheme}
    \end{equation}
    with $c_0,c_N$ a pair of hyperparameters that must be optimized such that (a) DSA is able to tunnel through high cost function barriers at the initial steps and (b) it is also able to converge and remain in a local minima at the final steps.

    \item The budget constraint scheme $\lambda_B(s)$, which we introduce for the first time in this work. If $\lambda_B$ were fixed, then the Markov chain would either be trapped inside local minima of the function $\lambda_B(\sum_i \omega_i -1)^2$ for large values of $\lambda_B$ or not enforce the budget constraint for small $\lambda_B$. Therefore, it becomes necessary to define an scheme for the budget constraint strength, for which a linear scheme $\lambda_B(s) = \lambda_B(0) + m_B\cdot s$, with $m_B > 0$, we have observed provides the best results.

    We define the initial and final budget constraint strengths as
    \begin{equation}
        \begin{split}
            \lambda_B(0) =&\ d_0 \frac{P_0^2}{\langle p^2\rangle}\cdot \textrm{median} \left\lbrace \big| \Delta_{i,a} \big|\right\rbrace  \\
            \lambda_B(N_s) =&\ d_N \frac{P_0^2}{\langle p^2\rangle}\cdot \textrm{median} \left\lbrace \big| \Delta_{i,a} \big|\right\rbrace ,
        \end{split}
        \label{eq:constraint_scheme}
    \end{equation}
    with $d_0,d_N$ two hyperparameters and $\langle p^2\rangle$ the mean squared asset price. Here we are assuming that the variations in the quadratic budget constraint $\left( \sum_{i} \omega_i - 1 \right)^2$ are of order $\langle p^2\rangle /P_0^2$ and remain around the same order of magnitude all along the Markov chain.

    \item The initial state preparation $\vec{\omega}(s=0)$. We use two methods: (i) \textit{uniform starting}, where we draw $\vec{\omega}(0)$ from an uniform distribution over the whole portfolio configuration space, and (ii) \textit{warm starting}, where we draw initial portfolios close to a target portfolio, which we take as the optimal continuous portfolio $\vec{\omega}_{\textrm{opt}}^c$. The distribution for the latter is a discrete multivariate normal distribution that is centered around the target portfolio and has constant variance $\sigma^2$,
    \begin{equation}
        p(n_i;n_i^c,\sigma) \sim
        \begin{cases}
            \exp\left[-\frac{(n_i-n_i^c)^2}{2\sigma^2}\right] & \quad \textrm{if}\quad 0 \leq n_i \leq n_i^{\max} \\
            0 & \quad \textrm{otherwise}
        \end{cases},
        \label{eq:warm_start}
    \end{equation}
    with $n_i^c \in \mathbb{R}$ the allocation of asset $i$ in the optimal continuous portfolio and $n_i \in \mathbb{N}$. If the continuous allocation for asset $i$ is $n_i^c=0$, then we set $n_i=0$ without drawing from the distribution.

    Warm starting the initial portfolios proves to be the optimal strategy for finding portfolios that maximize the quadratic utility, as the density of discrete portfolios around the continuous optimal portfolio scales as $P_0^{-1}$. Thus, the probability that the optimal discrete portfolio allocation lays close to the continuous one grows as the budget increases. Warm starting has already been shown to considerably speed up the solution finding process~\cite{Venturelli2019}.
\end{itemize}

As mentioned above, it might be the case while searching for optimal portfolios that the sum of weights $\sum_i \omega_i = \sum_i n_i \frac{p_i}{P_0}$ is not exactly 1, although that portfolio can still be nearly optimal. Thus, we allow for a relaxation of the budget constraint where we look for discrete portfolios that maximize the quadratic utility and their sum of weights is bounded by $1-\varepsilon \leq \sum_i \omega_i \leq 1$. Here $\varepsilon \geq 0$ is a parameter that indicates the maximal percentage of the portfolio that remains in cash (not allocated). In this work we set it to $\varepsilon = \frac{\langle p\rangle}{P_0}$, which ranges from $1\% - 0.0001\%$ in the problems studied in this work.

\subsection{Hyperparameter optimization}

DSA is very sensitive to the choice of hyperparameters, so we must run an hyperparameter optimization before we can look for optimal portfolios. For \textit{uniform starting} we have observed that the estimated number of steps needed to obtain optimal portfolios grows as $N_s \sim 10\cdot P_0$. Setting that as the number of steps, we observed that the initial parameters can be set to $(c_0=1,d_0=0)$, independent of $(c_N,d_N)$, and we use those hyperparameters for the rest of this work. To find the optimal values of $(c_N,d_N)$ for uniform starting, we use a 2d-grid algorithm that runs $n=3.000$ independent DSA computations with every possible set $(c_N,d_N)$ and then choose the set that maximizes the number of final portfolios with a quality distance below some threshold $d(\vec{\omega},\vec{\omega}^0) \leq d_t$. For \textit{warm starting} we fix $(c_N,d_N)$ to the values found for uniform starting and optimize $(c_0,d_0)$ using a 2d-grid search algorithm.

\begin{figure}[t]
    \centering
    \includegraphics[width=\linewidth]{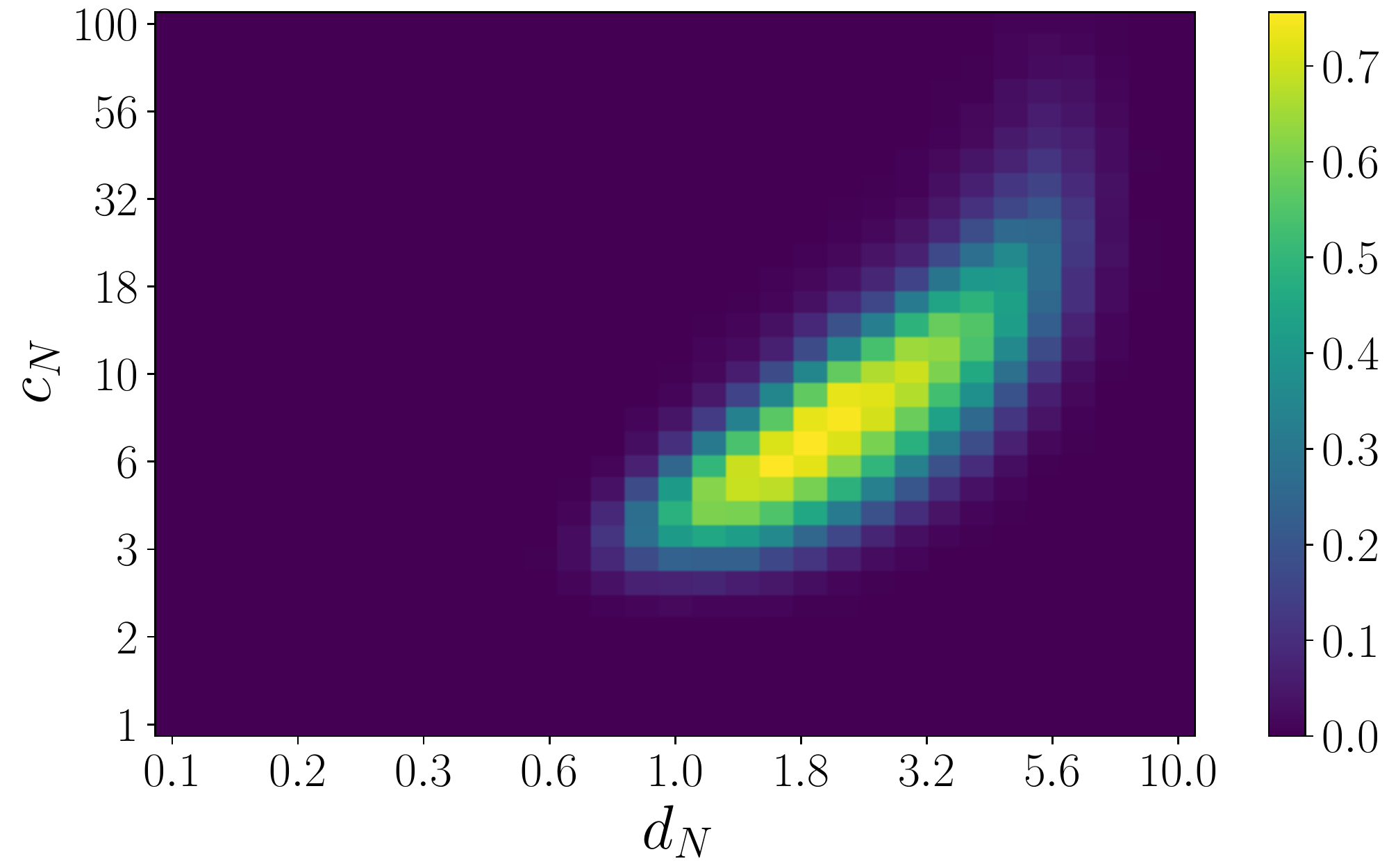}
    \caption{2d-grid search for the optimal hyperparameters $(c_N,\,d_N)$ with uniform starting. Each point shows the proportion of portfolios computed using DSA with a quality distance below $d(\vec{\omega}) \leq 10^{-8}$ among $n=3.000$ independent samples.}
    \label{fig:2d_grid_search}
\end{figure}

Fig.~\ref{fig:2d_grid_search} shows an example of the grid algorithm for finding the optimal set $(c_N,d_N)$ with uniform starting. Here the asset space consists of the first $L=100$ assets from the S\&P-500 index in alphabetical order. The expected returns and covariance matrix have been computed with the PyPortfolioOpt package~\cite{Martin2021} using historical daily prices data from 01/01/2008 to 31/12/2015. The risk aversion value is set to $\lambda=50$, the total portfolio budget is $P_0=10^5\$$ and there are no transaction costs $t_l=t_f=0$. The number of possible asset allocations is $\sim 2^{1060}$, making an exhaustive search for the global optimum portfolio unfeasible with every state of the art computational resource. At each point of the grid we run $n=3.000$ independent DSA instances with $N_s=3\cdot 10^6$ steps and show the proportion of portfolios after DSA that have a quality distance below $d(\vec{\omega}) \leq 10^{-8}$.

\section{Benchmarking the convex portfolio optimization problem}
\label{sec:benchmarking}

\begin{figure*}[t]
    \centering
    \includegraphics[width=\linewidth]{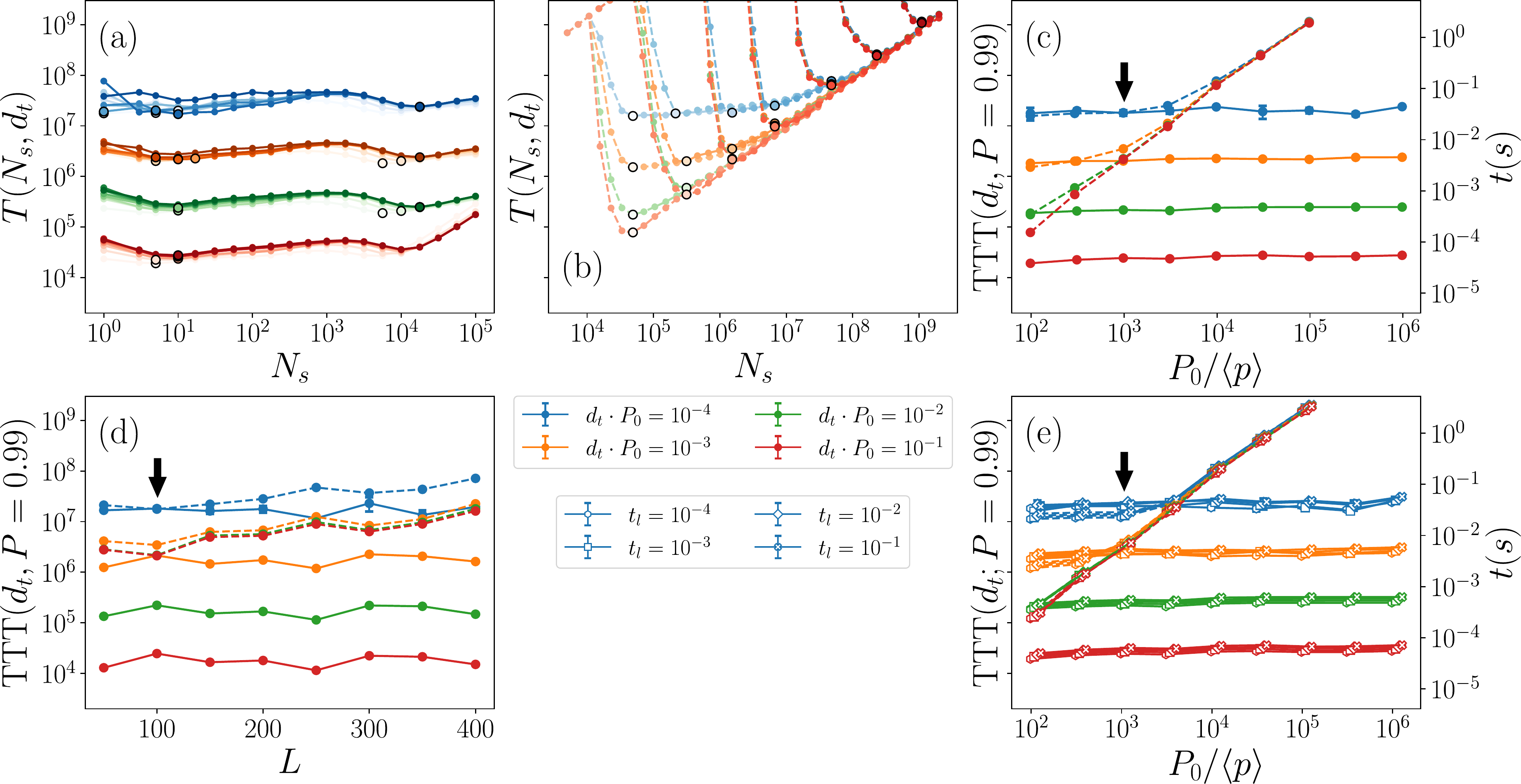}
    \caption{(a,b) Estimated run times of the DSA algorithm with warm starting (a) and uniform starting (b) to find a portfolio with distance below $d_t$ (see legend below (b)) with a probability $P=0.99$ against the number of steps $N_s$ and for different budgets $P_0$ (lighter colors for smaller budgets). The number of assets is $L=100$ and the risk factor is $\lambda=50$. Circles with black edges show the minimum estimated total time for each $d_t$ and $P_0$. (c) TTT extrapolated from the results of (a,b) against the portfolio budget $P_0$ rescaled by the mean asset price $\langle p\rangle$. Colors represent different distance targets $d_t$ scaled by the portfolio budget. We use warm starting (solid lines) in the DSA computations or uniform start (dashed lines). (d) TTT for a constant budget $P_0=10^5\$$ against different numbers of assets $L$. Colors and line types are equivalent to (c). The black arrows in (c,d,e) show the set of $(L,P_0)$ at which these panels coincide. (e) TTT with linear transaction costs (different markers) and varying budgets $P_0$. Colors and line types are equivalent to panel (c). We have shifted the lines with different $t_l$ a constant amount for better visual appreciation. The right y-axes in (c,e) show the estimated asymptotical run time of DSA in seconds for large $N_s \gg 1$.}
    \label{fig:ttt}
\end{figure*}

The standard metric for benchmarking combinatorial optimization algorithms is the \textit{time to solution}~\cite{Ronnow2014}, which estimates the time taken by the algorithm to find the global optimum with some fixed probability. Because the search space of portfolio optimization problems grows exponentially with the number of assets $L$ and budget $P_0$, we cannot find the global optimum portfolio in a reasonable time, which is needed to estimate the time to solution. Thus, we propose to measure the \textit{time to target} (TTT) instead as a proper benchmark for the discrete portfolio optimization problem. This benchmark, which we will define below, estimates the time taken by the algorithm to find a local optimum portfolio with quality distance below some target $d(\vec{\omega}) \leq d_t$ with some probability $P$ that we fix to $P=0.99$ for the rest of this work.

\subsection{Estimated DSA run times}

We first estimate the required run time $T(N_s,d_t)$ of DSA to obtain a portfolio that satisfies $d(\vec{\omega}) \leq d_t$ with a fixed number of steps $N_s$. Because of the stochastic nature of DSA, we compute $n$ independent runs for the same problem. Then, we can estimate the probability $p(N_s,d_t)$ that a single DSA simulation obtains a portfolio with distance below $d_t$ by counting the number of portfolios that satisfy that condition. Using that probability we estimate the number of independent runs $R(N_s,d_t)$ that is required to get at least one portfolio with distance below $d_t$ with a probability $P=0.99$
\begin{equation}
    (1-P) = \left(1 - p(N_s,d_t)\right)^{R(N_s,d_t)}
\end{equation}
and define the required run time as
\begin{equation}
    T(N_s,d_t) = N_s\cdot R(N_s,d_t).
\end{equation}

We show in Figs.~\ref{fig:ttt}(a,b) the estimated DSA run time $T(N_s,d_t;P_0)$ against different numbers of steps $N_s$, budgets $P_0=10^4,3\cdot 10^4,10^5,\dots,10^8\,\$$ (color graduation) and distance targets $d_t$ (different colors, see legend below (b)). In this portfolio optimization problem we take the first $L=100$ assets of the S\&P-500 Index in alphabetical order and compute the expected returns and correlation matrix using historical price data from 01/01/2008 to 31/12/2015. The risk balance factor is $\lambda=50$ and there are no transaction costs to $t_l=t_f=0$. In Fig.~\ref{fig:ttt}(a) we use warm starting, while in (b) we use uniform starting. We observe that the dependence of the run times on $N_s$ is non trivial in both cases and there is always some finite number of steps for which $T(N_s,d_t)$ (we omit $P_0$ for simplicity) is minimal, which we mark using circles with black edges for each $P_0$ and $d_t$. For warm starting the optimal number of steps is of order $N_s \sim 10$ (in some cases around $10^4$), indicating that starting DSA with a massive number of independent runs (here $n = 5\cdot 10^7$) is almost guaranteed to fall very close to the global optimum at least once.

\subsection{Time to target against the number of assets and budget}

We define the TTT of DSA as the optimal run time
\begin{equation}
    \textrm{TTT}(d_t) = \min_{N_s} \left\lbrace T(N_s,d_t) \right\rbrace.
\end{equation}
In Figs.~\ref{fig:ttt}(c,d) we show the estimated TTT against different distance targets $d_t$ (different colors). In (c) we use the same portfolio optimization problem of panels (a,b), while in (d) we have different numbers of assets $L$ (taken alphabetically from the S\&P-500 Index) and a fixed budget $P_0=10^5\,\$$. Dashed lines show the estimated TTT for uniform starting and solid lines correspond to warm starting. The error bars mark one estimated standard deviation of the TTT.

We show in Tab.~\ref{tab:ttt} the extrapolated asymptotical TTT scaling for large $L$ or $P_0$ for uniform and warm starting. We note that all extrapolated scalings are polynomial in either $L$ or $P_0$. Thus, the portfolio optimization problem seems to be not 'hard' to solve using classical resources. Moreover, our data suggest that the TTT remains almost constant with $P_0$ when using warm starting, meaning that massively sampling around the optimal continuous solution is almost always guaranteed to find the global optimum.

\begin{table}[b!]
\centering
\begin{tabular}{| c | c |}
 \hline
  &  \textrm{Fixed} $L$ \\
 \hline
 \textrm{Unif. start} & $\left(P_0\right)^{m},\ m = 1.38\pm 0.02$  \\ [0.6ex]
 \hline
 \textrm{Warm start} & $\left(d_t\right)^{-1} \cdot \left(P_0\right)^{m},\ m = 0.05\pm 0.02$  \\ [0.6ex]
 \hline
 \hline
 & \textrm{Fixed} $P_0$  \\
 \hline
 \textrm{Unif. start} & $10^{m L},\ m = 7\cdot 10^{-4} \pm 3\cdot 10^{-4}$  \\ [0.6ex]
 \hline
 \textrm{Warm start} & $10^{m L},\ m = 1.6\cdot 10^{-3} \pm 3\cdot 10^{-4}$ \\ [0.6ex]
 \hline
\end{tabular}
\caption{Estimated asymptotical scaling of the TTT with $L$ and $P_0$ using uniform and warm starting. The $m$ parameters have been computed with a least squares fit of the estimated TTTs, where we also provide the estimated standard deviation.}
\label{tab:ttt}
\end{table}

In the secondary y-axes of Figs.~\ref{fig:ttt}(c,e) we show the estimated run time of a single DSA computation in seconds. The scaling from number of steps to seconds was fitted using the median run time in seconds of $n=512$ independent runs with uniform starting, numbers of assets $L=100,\,200$ and initial budgets $P_0=10^4, 10^5, 10^6 \$$ using $5$ samples for each set $(L,P_0)$. The computations were done using an \texttt{AMD Ryzen threadripper pro 3955wx 16-coresx32} processor.

\subsection{Time to target with linear costs}

To explore the scaling of DSA in more general convex portfolio optimization problems we now analyze the same problem of Figs.~\ref{fig:ttt}(a,b,c) with non zero linear costs $t_l > 0$ and without fixed costs $t_f=0$, for which the continuous problem is still convex. For this problem we optimize a portfolio at $01/12/2015$ using DSA and use it as a past portfolio $\vec{\omega}^0$ that we rebalance at $01/01/2016$ with linear transaction costs using DSA. We show in Fig.~\ref{fig:ttt}(e) the estimated TTT against the total budget, distance target and linear costs $t_l=10^{-4},10^{-3},10^{-2},10^{-1}$ (different markers, see legend below Fig.~\ref{fig:ttt}(b)). Line types and colors are the same as in Figs.~\ref{fig:ttt}(c,d). We observe the same asymptotical scaling of the TTT for both warm and uniform starting as when there are no linear costs.

\begin{figure*}[ht]
    \centering
    \includegraphics[width=\linewidth]{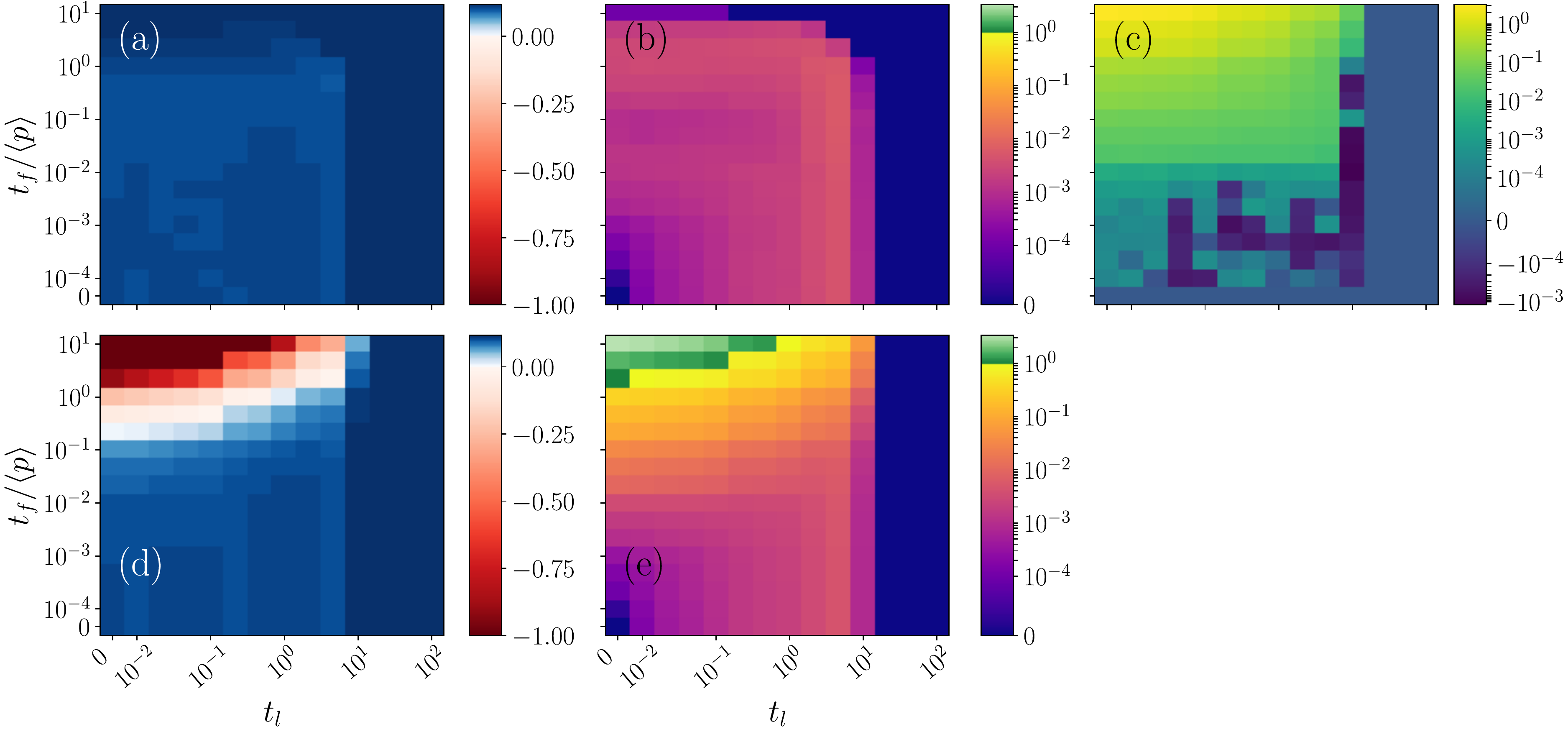}
    \caption{(a,d) Average yearly return of portfolios that have been monthly rebalanced with DSA from January until December against different fixed $t_f$ and linear $t_l$ transaction costs. The average is taken over 10 years and with 8 different initial portfolios per year. In (a) the DSA algorithm includes fixed transaction costs in the cost function, while (d) does not. (b,e) Average paid transaction costs over one year of the portfolios in (a,d). In (b) the DSA algorithm includes fixed transaction costs in the cost function, while (e) does not. (c) Difference between the quadratic utilities with transaction costs $Q_t$ when DSA incorporates fixed costs in the cost function and when it does not.}
    \label{fig:tcosts}
\end{figure*}

The asymptotical scaling extrapolations for the TTT in this Section suggest that the portfolio optimization problem can be solved in a time proportional to the total budget $P_0$ using DSA with uniform starting, while it can be solved in near constant time with DSA with warm starting. In the case when there are linear costs the scalings are the same, suggesting that the convex portfolio optimization problem can be solved in under polynomial time using only classical resources.

\section{Non convex optimization and out of sample results}
\label{sec:non-convex}

When the continuous portfolio optimization problem of Eq.~\eqref{eq:quadratic_utility_costs} becomes non convex, e.g. by having fixed transaction costs $t_f \neq 0$, we cannot guarantee to find the optimal continuous asset allocation in polynomial time. In this case the quality distance $d(\vec{\omega})$ is no longer a good metric and doing a TTT benchmark becomes unfeasible. However, in the context of non convex quadratic utilities we can still show the advantage that using the DSA algorithm can have over other solvers that only optimize convex problems. We will show this by comparing the results of two optimizations, one of which incorporates fixed costs in the cost function, which we will refer to in this Section as \textit{general} DSA, while the other only includes linear costs and optimizes a purely convex function, we call it \textit{convex-only} DSA.

A problem of interest in the financial industry is multi period portfolio optimization (MPO)~\cite{Boyd2017}. Because the asset prices, expected returns and correlations change over time, the portfolio needs to be rebalanced periodically so that the portfolio always aims at maximum returns over risk. When assets are traded at each rebalancing this incurs into transaction fees, which must be taken into account to keep the returns at an optimal value.

The MPO problem that we solve in this work consists in keeping over one year a portfolio that is monthly rebalanced. The initial portfolio is one that is nearly optimal at $01/12/y$ of the previous year. We compute this MPO problem for 10 different years, $y=2012,2013...,2021$, and with $10$ different initial portfolios for each year. To be able to compare the portfolios between different years $y$ we set the initial budget at $P_0(y) = 10^3 \cdot \langle p(y)\rangle$, with $\langle p(y)\rangle$ the mean asset price at $01/01/y$, and scale the fixed transaction costs as $t_f(y) = r_f\cdot \langle p(y)\rangle$. The expected returns and correlations are computed using historical data from $01/01/2008$ until the corresponding date.

We show in Figs.\,\ref{fig:tcosts}(a,d) the average yearly returns over all years and initial portfolios of the above MPO problem using general DSA (a) and convex-only DSA (d). These returns are defined as
\begin{equation}
    \mu(y) = \frac{P_{01/01/y+1} - T_{c,y}}{P_0} - 1
\end{equation}
with $P_t$ the total portfolio value at time $t$, including leftover cash, $P_0$ the initial budget and $T_{c,y}$ the total transaction costs paid in year $y$. We observe in Fig.\,\ref{fig:tcosts}(d) that when using convex-only DSA the average portfolio return is negative for high fixed costs and low linear costs, as convex-only DSA is not able to avoid paying fixed transaction costs. If we use general DSA, the average return remains stable around $+12\%$, with a small relative increase in the returns where high transaction costs have prevented general DSA from rebalancing. While this might seem counter intuitive, this comes from the stochastic nature of market prices~\cite{Cho2011,Kolm2014,Kazak2019}, making the expected returns we are optimizing in the cost function only approximately valid. This introduces some degree of noise in the MPO problem that can make doing less rebalancing better than frequent rebalancing.

Figs.\,\ref{fig:tcosts}(b,e) show the average paid transaction costs rescaled by the initial budget $T_{c,y}/P_0$ using general DSA (b) and convex-only DSA (e). As expected, general DSA lowers the paid transaction costs over convex-only DSA. We observe hints of a phase transition in the portfolio rebalancing problem when linear costs $t_l$ are greater than $10$, as the average paid transaction costs go to zero beyond that boundary both in Figs.\,\ref{fig:tcosts}(b,e).

While the stochastic nature of the portfolio optimization problem can lower the total returns achieved when using real asset prices, we can still show the advantage of using general DSA over other only convex methods by comparing the total quadratic utility with transaction costs over the whole rebalancing period
\begin{equation}
\begin{split}
    Q_R = \sum_{t=1}^{12} &\ \vec{\mu}(t)^T \vec{\omega}(t) - \frac{\lambda}{2} \vec{\omega}(t)^T S(t) \vec{\omega}(t) \\
    & - \frac{t_f}{P_0} \sum_{i} \big[ 1 - \delta(\omega_i(t) - \omega_i(t-1)) \big] \\
    & + t_l \sum_{i} |\omega_i(t) - \omega_i(t-1)|,
\end{split}
\end{equation}
with $\vec{\omega}(t)$ the rebalanced portfolio at month $t$, $\vec{\omega}(0)$ the initial portfolio at $01/12/y-1$ and $P_0$ the initial budget. We show in Fig.\,\ref{fig:tcosts}(c) the difference $Q_R^{\textrm{gen}} - Q_R^{\textrm{conv-only}}$ and observe that, when fixed costs are bigger than linear costs, the total quadratic utility is almost always bigger using general DSA. While for some transaction fees $(t_f,t_l)$ the difference is negative, these differences are relatively small and might come from the stochastic nature of the portfolio optimization problem.

\section{Conclusions}
\label{sec:conclusions}

In this work we describe a DSA algorithm that avoids the mapping to binary variables usual in combinatorial optimization problems and allows us to solve the portfolio optimization problem with fixed and linear transaction costs in an exact form. We benchmark DSA for solving this problem by estimating the time to target, a measure of the computational run time needed to achieve an optimal portfolio with a given quality of solution. From this estimation we extrapolate a polynomial asymptotic scaling of the run time of DSA with the budget's size and number of assets when DSA is started at random from an uniform distribution. If we use warm starting, then we have shown that the run time remains roughly constant. We also observe that the asymptotical scalings are similar when there are linear transaction costs in the cost function. Therefore, our computations suggest that the portfolio optimization problem with only convex functions in the quadratic utility is not 'hard' to solve using classical resources.

Apart from benchmarking the convex portfolio optimization problem, we have tested the DSA algorithm for solving the multi period portfolio optimization problem with non convex transaction costs using historical asset prices. We show how DSA is sensible to non convex functions in the quadratic utility and is able to avoid fixed and linear transaction costs, thus making this algorithm very suitable for application in the portfolio optimization problem with other kinds of transaction costs or constraints.

\section{Acknowledgments}
We acknowledge the CSIC Interdisciplinary Thematic Platform (PTI+) on Quantum Technologies (PTI-QTEP+). This research is part of the CSIC program for the Spanish Recovery, Transformation and Resilience Plan funded by the Recovery and Resilience Facility of the European Union, established by the Regulation (EU) 2020/2094.
We also acknowledge support by the Spanish project PGC2018-094792-B100 (MCIU/AEI/FEDER, EU). The results and analysis presented in this paper were possible thanks to the access granted to computing resources at the Galicia Supercomputing Center, CESGA, including access to FinisTerrae II/III.
\bibliography{bibi}

\end{document}